\documentclass[,final]{aipproc}

\layoutstyle{6x9}

\begin{document}

\title{Cosmological constraints for a two brane-world system with single equation of state}


\author{Juan L. P\'erez}{
  address={Divisi\'on de Ciencias e Ingenier\'ias campus Le\'on, Universidad de Guanajuato, Guanajuato, M\'exico}
}

\author{Rub\'en Cordero}{
  address={Departamento de F\'{\i}sica, Escuela Superior de
F\'{\i}sica y Matem\'aticas del IPN\\ Unidad Adolfo
L\'opez Mateos, Edificio 9, 07738, M\'exico D.F., M\'exico}
}

\author{L. Arturo Ure\~na-L\'opez}{
  address={Divisi\'on de Ciencias e Ingenier\'ias campus Le\'on,
    Universidad de Guanajuato, Guanajuato, M\'exico}
}

\begin{abstract}
We present the study of two 3-brane system embedded in a 5-dimensional
space-time in which the fifth dimension is compactified on a
$S^{1}/Z_{2}$ orbifold. Assuming isotropic, homogeneous, and static
branes, it can be shown that the dynamics of one brane is dominated
by the other one when the metric coefficients have a particular
form. We study the resulting cosmologies when one brane is dominated
by a given single-fluid component.
\end{abstract}

\maketitle

\section{Introduction}
In the last years, there has been a great expectation about braneworld
scenario as an alternative proposal to explain some unsolved problems in particle physics and cosmology, including the hierarchy
and the dark matter-energy problems. It is in this context
that diverse braneworld models have emerged in an attempt to put a final point
to these problems.

Superstring and M-theory suggest that we may live in a world that has
more than three spatial dimensions \cite{Schwarz:2008kd}. Because only
three of these dimensions are presently observable, one has to explain why the
others are hidden from detection. One such explanation is the
so-called Kaluza-Klein (KK) compactification (see \cite{k-k-1}
and \cite{k-k-2} for a review), according to which the extra dimensions
are very small, probably with a size of the order of the Planck
length. As a consequence, modes that have momentum in the directions
of the extra dimensions are excited at currently inaccessible
energies.

In 1998, N. Arkani-Hamed,  S. Dimopoulus and G. R. Dvali (ADD) \cite{ADD-1} pointed out
that the extra dimensions are not necessarily small, and
may even be in the scale of millimeters. This model assumes that the
Standard Model fields are confined to a three dimensional surface (a
3-brane) embedded in a larger dimensional background spacetime (bulk) where the gravitational
field is free to propagate. Additional fields may
live only on the brane or in the bulk, provided that their
current undetectability is consistent with experimental
bounds \cite{Hoyle:2004cw}.

An alternative approach was proposed by L. Randall and R. Sundrum
(RS) \cite{Randall:1999ee}, which will be hereafter referred to as RS1
model (see \cite{Wang:2006ue} for a general treatment of two 3-branes
in a RS setup). The bulk in this model is 5-dimensional, with the
extra dimension being compactified on an $S_{1}/Z_{2}$ orbifold i.e.
the extra dimension $y$ is periodic, and its ends points are
identified. In such a setting, the bulk necessarily contains two
3-branes, located, respectively, at the fixed points $y = 0$, and
$y=y_{c}$. The brane at $y = 0$ is usually called hidden (or Planck)
brane, and the one at $y = y_{c}$ is called visible (or TeV) brane.

It is then natural to ask if the evolution of one brane is related to
the other one (just as in the RS1 model but in general for any
metric). Binetruy et al \cite{Binetruy:1999ut} have showed
that there exists a relationship between the energy density for the two
branes. The goal of this paper is to generalize these relations found by
Binetruy et al, and to analyze their cosmological evolution. As an example, we examine a single fluid component for the brane universes in this paper.

\section{Mathematical background}
We begin with the most general 5-dimensional metric in which flat
branes lie in a homogeneous and isotropic subspace (located at $y=0$
and at $y=y_{c}$)
\begin{equation}
  \label{metrica}
  ds^{2} =-n^{2}(\tau,y) dt^{2} + a^{2}(\tau,y) \delta_{i j} dx^{i}
  dx^{j} + b^{2}(\tau,y) dy^{2} \, .	
\end{equation}
The imposed symmetries:
$(x^{\mu},y)\rightarrow(x^{\mu},-y)$ (reflection) and
$(x^{\mu},y)\rightarrow(x^{\mu},y+2my_{c})\ m=1,2,...\ $ (compactification), demand that
each one of the metric coefficients, $a(t,y)$, $n(t,y)$, and $b(t,y)$, is subjected to the following conditions \cite{Binetruy:1999ut} :
\begin{eqnarray}
  F(t,y)&=&F(t,\left|y\right|) \, , \\
  \label{condicion1}	
  \left[F'\right]_{0}&=&2F'|_{y=0+} \, , \\
  \label{condicion2}	
  \left[F'\right]_{c}&=&-2F'|_{y=y_{c}-} \, , \\
  \label{condicion3}	
  F''(t,y) &=& F"+\left[F'\right]_{0}\delta(y)+\left[F'\right]_{c}\delta(y-y_{c})
  \, . \label{condicion4}
\end{eqnarray}
In the above equations, $F(t,y)$ represents any of the metric
coefficients, the prime denotes derivative with respect to $y$, the square
brackets denote the \emph{jump} in the first derivative at $y=0$ and
at $y=y_{c}$. Eq. ~(\ref{condicion4}) is obtained if we demand
that $|y|^\prime =1$, and $\left|y\right|^{\prime
  \prime} = 2\delta(y) - 2\delta(y-y_{c})$ in the interval
$[0,y_{c}]$. We define  $F" \equiv
\frac{d^{2}F(t,\left|y\right|)}{d\left|y\right|^{2}}$. It is necessary
here to recall that the subindex $0$ will be used for quantities
valued at $y=0$; likewise, the subindex $c$ will be used for quantities
valued at $y=y_{c}$.

The five-dimensional Einstein equations, $\tilde{G}_{AB} =
\kappa^{2}_{(5)} \tilde{T}_{AB}$ for metric~(\ref{metrica}) are
\begin{eqnarray}
  \tilde{G}_{00} &=& 3\left\{\frac{\dot{a}}{a} \left(
      \frac{\dot{a}}{a} + \frac{\dot{b}}{b} \right) -
    \frac{n^{2}}{b^{2}} \left[ \frac{a''}{a} + \frac{a'}{a} \left(
        \frac{a'}{a} - \frac{b'}{b} \right) \right] \right\} \,
  , \label{einstein1} \\
  \tilde{G}_{ij} &=& \frac{a^{2}}{b^{2}} \delta_{ij} \left\{
    \frac{a'}{a} \left( \frac{a'}{a} + 2 \frac{n'}{n} \right) -
    \frac{b'}{b} \left( \frac{n'}{n} + 2\frac{a'}{a} \right) + 2
    \frac{a''}{a} + \frac{n''}{n} \right\} \\
  && + \frac{a^{2}}{b^{2}} \delta_{ij} \left\{ \frac{\dot{a}}{a}
    \left( -\frac{\dot{a}}{a} + 2\frac{\dot{n}}{n} \right) - 2
    \frac{\ddot{a}}{a} + \frac{\dot{b}}{b} \left( -2\frac{\dot{a}}{a}
      + \frac{\dot{n}}{n} \right) - \frac{\ddot{b}}{b} \right \} \,
  , \label{einstein2} \\
  \tilde{G}_{05} &=& 3\left( \frac{\dot{a}}{a} \frac{n'}{n} +
    \frac{\dot{b}}{b} \frac{a'}{a} - \frac{\dot{a}'}{a} \right) \,
  , \label{einstein3} \\
  \tilde{G}_{55} &=& 3 \left\{ \frac{a'}{a} \left( \frac{a'}{a} +
      \frac{n'}{n} \right) - \frac{b^{2}}{n^{2}} \left[
      \frac{\ddot{a}}{a} + \frac{\dot{a}}{a} \left( \frac{\dot{a}}{a}
        - \frac{\dot{n}}{n}\right) \right] \right\} \, .
\end{eqnarray}
For the moment we consider the energy-momentum tensor associated with the brane in an empty bulk
\begin{equation}
  \label{tme}
  \tilde{T}^{A}_{B} =
  \frac{\delta(y)}{b_{0}} diag(-\rho_{0},p_{0},p_{0},p_{0},0) +
  \frac{\delta(y-y_{0})}{b_{c}} diag(-\rho_{c},p_{c},p_{c},p_{c},0) \,
  ,
\end{equation}
i.e., only the brane contributions to the energy-momentum tensor at
$y=0$ and $y=y_{c}$ are taken into account, this leads us, using the
Bianchi identity, $\nabla_{A}\tilde{G}^{A}_{B} = 0$, to an equation of
conservation for the energy density of the form
\begin{equation}
  \label{conservacion}
  \dot{\rho_{0}} + 3(p_{0}+\rho_{0}) \frac{\dot{a}_{0}}{a_{0}} = 0 \, .
\end{equation}

According to the Israel \textit{junction conditions}, we need to
describe the presence of an energy density in terms of a discontinuity
in the metric across the origin in the extra coordinate. In this way, using
Eqs. ~(\ref{condicion4}), and (\ref{tme}) in the $(0,0)$
and $(i,j)$ components of the Einstein tensor valued at $y=0$, it is not
difficult to find the following relations for the metric coefficients:
\begin{equation}
  \frac{\left[a'\right]_{0}}{a_{0}b_{0}} =
  -\frac{\kappa^{2}_{(5)}}{3}\rho_{0} \, , \quad
  \frac{\left[n'\right]_{0}}{n_{0}b_{0}} =
  \frac{\kappa^{2}_{(5)}}{3} (3p_{0} + 2\rho_{0}) \, . \label{salto2}
\end{equation}
This implies that the \emph{jump} in the first derivative of the metric
coefficients is proportional to the energy density across the
origin. Similarly, at $y=y_{c}$, we have
\begin{equation}
\frac{\left[a'\right]_{c}}{a_{c}b_{c}} =
-\frac{\kappa^{2}_{(5)}}{3}\rho_{c} \, , \quad
\frac{\left[n'\right]_{c}}{n_{c}b_{c}} = \frac{\kappa^{2}_{(5)}}{3}
(3p_{c} + 2\rho_{c}) \, . \label{salto4}
\end{equation}

To be consistent with the Friedmann-Robertson-Walker (FRW) metric at
the origin, we demand that $n_{0}=n(t,y=0)=1$, and that
$\dot{n}_{0}=0$. With the aid of Eqs. ~(\ref{salto2}),
and~(\ref{salto4}), in the $(5,5)$ component of Einstein tensor, together
with the conditions (\ref{condicion2}) and~(\ref{condicion3}), we
obtain the modified Friedmann equation,
\begin{equation}
  \label{friedmann}
  \frac{\dot{a}^{2}_{0}}{a^{2}_{0}} + \frac{\ddot{a}_{0}}{a_{0}} =
  -\frac{\kappa^{4}_{(5)}}{36} \rho_{0} (\rho_{0} + 3p_{0}) \, .
\end{equation}

For a singe component dominated universe with an equation of state
$p_{0}=\omega_{0}\rho_{0}$, we get
\begin{equation}
  \label{densidad}
  \kappa^{2}_{(5)}\rho_{0}(t) = \frac{6}{(3+3\omega_{0})t} \, ,
\end{equation}
which is valid only for $\omega \neq -1$, whereas for $\omega=-1$ the
energy density remains constant in time. The last two equations are
fundamentally different from the standard Friedmann equation in that the
latter depends linearly in the energy density, whereas in
Eq. ~(\ref{friedmann}) the dependence is on the \emph{square} of
the energy density.

\section{Connection between the two branes}
Let us begin with an ansatz for the metric coefficient
$a(t,\left|y\right|)$ such that it depends on two time dependent
parameters, $\lambda (t)$ and $\alpha(t)$, as
\begin{equation}
  \label{forma1}
  a(t,\left|y\right|) = \alpha f(\lambda\left|y\right|) \, .
\end{equation}
Function $f$, and its derivative $f'$, are well defined at $[0,y_{c}]$,
actually $f(0)=1$. In this way,
\begin{equation}
  \label{forma2}
  a(t,\left|y\right|) = a_{0} f(\lambda\left|y\right|) \, .
\end{equation}
According to Eqs. ~(\ref{condicion2}) and~(\ref{condicion3}), we
obtain
\begin{equation}
  \left[a'\right]_{0} = 2\lambda a_{0}f'_{0}, \quad
  \left[a'\right]_{c} = -2\lambda a_{0}f'_{c} \, , \label{salto6}
\end{equation}
where $f^\prime \equiv
\frac{df(\lambda\left|y\right|)}{d(\lambda\left|y\right|)}$. From
Eqs. ~(\ref{salto2}), (\ref{salto4}), and~(\ref{salto6}), we find
\begin{equation}
\label{restriccion1}
\kappa^{2}_{(5)}\rho_{c}=-\frac{b_{0}f'_{c}}{b_{c}f'_{0}f_{c}}\kappa^{2}_{(5)}\rho_{0}.
\end{equation}
In the last equation, $f_{c}$ and $f^\prime_{c}$ are functions of
$\lambda$, and consequently, functions of $\rho_{0}$, namely,
\begin{equation}\label{lambda}
  \lambda = -\frac{b_{0}}{6f'_{0}} \kappa^{2}_{(5)} \rho_{0} \, .
\end{equation}

These last two results show that there exists a connection between the
two branes just in the form of topological constraints. Now, we are interested in the connection between the equations of state in each brane. In analogy to Eq. (\ref{restriccion1}), but now for the ansatz
\begin{equation}~
  \label{forma3}
  n(t,\left|y\right|)=g(\beta\left|y\right|),
\end{equation}
where $g(0)=1$, and using Eqs. ~(\ref{salto2}),
and~(\ref{salto4}), we find
\begin{equation}
  \label{restriccion2}
  \kappa^{2}_{(5)}\rho_{c}(2+3\omega_{c})=-\frac{b_{0}g'_{c}}{b_{c}g'_{0}g_{c}}\kappa^{2}_{(5)}\rho_{0}(2+3\omega_{0})
  \, ,
\end{equation}
where $g'\equiv\frac{dg(\beta\left|y\right|)}{d(\beta\left|y\right|)}$.
In the former equation, both $g_{c}$ and $g'_{c}$ are functions of $\beta$, where
\begin{equation}
  \label{beta}
  \beta=\frac{b_{0}}{6g'_{0}}\kappa^{2}_{(5)}\rho_{0}(2+3\omega_{0})
  \, .
\end{equation}
Combining Eq. ~(\ref{restriccion2}) with
Eq. ~(\ref{restriccion1}), we find that $\omega_{c}$ is connected
to $\omega_{0}$ and $\rho_{0}$, namely,
\begin{equation}
\label{ecuaciondeestado}
(2+3\omega_{c})=\frac{f_{c}}{f'_{c}}\frac{g'_{c}}{g_{c}}\frac{f'_{0}}{g'_{0}}(2+3\omega_{0}))
\, .
\end{equation}
This last equation gives us the relationship between the equations of
state valued at the position of each
brane. Eqs. ~(\ref{restriccion1}), and~(\ref{ecuaciondeestado}),
are the main results of this paper.

As an example, let us take $b=1$, and
\begin{equation}
a(t,\left|y\right|) = a_{0}(t)e^{\lambda\left|y\right|} \, , \quad
n(t,\left|y\right|)= e^{\beta\left|y\right|} \, . \label{RS2}
\end{equation}
 We can see that $f'_{0}=g'_{0}=1$, $f_{c}=f'_{c}$ and $g_{c}=g'_{c}$,
and therefore $\omega_{0}=\omega_{c}$. When $\lambda= \beta < 0$, which corresponds to Anti de Sitter bulk, we have $\omega_0=-1$ and it is precisely the case of the RS setup \cite{Randall:1999ee}.

\section{Conclusions}
In this work we have showed that, in a two-brane setup, there exists a
relationship between the equation of state of the two brane universes. We considered a $S_{1}/Z_{2}$ compactification,
and the metric was such that we recover the FRW setup at the location
$y=y_{0}$. Our results are of general application, and  we were able
to recover the Randall-Sundrum cosmology under simple assumptions.

More involved models can be analyzed, in which the dynamics of the TeV
brane can match that of the standard cosmology. This is work under
progress that we expect to present elsewhere.


\begin{theacknowledgments}
We are grateful to the Divisi\'on de Gravitaci\'on y F\'isica
Matem\'atica (DGFM) for the opportunity to present this work at the
VIII Taller of the DGFM. R. C. was partially supported by SNI-M\'exico, CONACyT research grant J1-60621-I, COFAA-IPN and SIP-IPN grant 20111070.
\end{theacknowledgments}

\bibliographystyle{aipproc}

\bibliography{BIBLIOGRAFIA-BRANAS}

\end{document}